\documentclass[11pt,final,reqno]{amsart}
\newcommand{\normalspacing}{\renewcommand{\baselinestretch}{1.1}\tiny\normalsize}

\normalspacing

\usepackage{amsmath,amsfonts,amssymb,alltt,verbatim,xspace}

\usepackage[agsm]{harvard} % options here: agsm, jphysicsB
\usepackage[final]{graphicx}

\theoremstyle{theorem}
\newtheorem{thm}{Theorem}
\newtheorem{ass}{Assumption}

\newcommand{\bQ}{\mathbf{Q}}

\newcommand{\br}{\hat{\mathbf{r}}}
\newcommand{\bU}{\mathbf{U}}

\newcommand{\Div}{\nabla\cdot}

\newcommand{\ppt}[1]{\frac{\partial #1}{\partial t}}
\newcommand{\ppT}[1]{\frac{\partial #1}{\partial T}}

\newcommand{\ppz}[1]{\frac{\partial #1}{\partial z}}
\newcommand{\ppzz}[1]{\frac{\partial^2 #1}{\partial z^2}}

\newcommand{\grad}{\nabla}

\newcommand{\ip}[2]{\left<#1,#2\right>}

\newcommand{\vf}{\varphi}

\newcommand{\RR}{\mathbb{R}}

\newcommand{\BKL}{\textsc{\emph{BBL}}\xspace}

\title[On exact solutions and numerics for thermocoupled ice sheets]{On exact solutions and numerics for \\ cold, shallow, and thermocoupled ice sheets}

\author[Bueler and Brown]{Ed Bueler and Jed Brown}

\thanks{\hspace{-4mm}\textsc{DRAFT} date: \today. \\ \hspace{-5mm} Dept.~of Mathematics and Statistics, University of Alaska, Fairbanks AK 99775-6660; e-mail \texttt{ffelb\@@uaf.edu}.}

\begin{document}

\begin{abstract}  This three section report can be regarded as an extended appendix to \cite{BBL}.  First we give the detailed construction of an exact solution to a standard continuum model of a cold, shallow, and thermocoupled ice sheet.  The construction is by calculation of compensatory accumulation and heat source functions which make a chosen pair of functions for thickness and temperature into exact solutions of the coupled system.  The solution we construct here is ``Test \textbf{G}'' in \cite{BBL} and the steady state solution ``Test \textbf{F}'' is a special case.  In the second section we give a reference C implementation of these exact solutions.  In the last section we give an error analysis of a finite difference scheme for the temperature equation in the thermocoupled model.  The error analysis gives three results, first the correct form of the Courant-Friedrichs-Lewy (CFL) condition for stability of the advection scheme, second an equation for error growth which contributes to understanding the famous ``spokes'' of \cite{EISMINT00}, and third a convergence theorem under stringent fixed geometry and smoothness assumptions. \end{abstract}

\maketitle
\thispagestyle{empty}

\section{Derivation of the exact solution}

\subsection{Review: Equations of the continuum model}  The flat bed, frozen base case of the cold shallow ice approximation is, for the purposes of this paper, taken to be the equations in the ``Continuum Model'' part of \cite{BBL}; that paper is hereafter referred to as ``\BKL.''  The notation used in, the physics of, and the boundary conditions for the continuum model are all laid out in \BKL.  The equations are repeated here for convenience:
\begin{align}
\text{mass-balance:} &&    &\ppt{H} = M - \Div \bQ, \label{flow}\\
\text{temperature:} &&  &\ppt{T} = \frac{k}{\rho c_p} \ppzz{T} - \bU\cdot \grad T - w \ppz{T} + \Sigma. \label{temp} \\
\text{effective shear stress:} &&    &\ip{\sigma_{xz}}{\sigma_{yz}} = - \rho g (H-z) \grad H, \label{stress}\\
\text{constitutive function:} &&    &F(T,\sigma) = A \exp\left(\frac{-Q}{R T}\right) \sigma^{n-1},  \label{ArrGlenNye} \\
\text{horizontal velocity:} &&       &\bU = - 2 \rho g \grad H \int_0^z F(T,\sigma) (H-\zeta)\,d\zeta, \label{horvel}\\
\text{map-plane flux:} &&       &\bQ = \int_0^H \bU\,dz,\label{flux} \\
\text{vertical velocity:} &&       &w = -\int_0^z \Div \bU\,d\zeta, \label{incomp}
\end{align}
\begin{align}
\text{strain heating:} &&       &\Sigma = \frac{1}{\rho c_p} \ip{\sigma_{xz}}{\sigma_{yz}} \cdot \ppz{\bU}. \label{strainheat}
\end{align}
The primary scalar unknowns are the thickness $H$, which equals the surface elevation in this flat bed case, and the absolute temperature $T$.  Note $\sigma = \left(\sigma_{xz}^2 + \sigma_{yz}^2\right)^{1/2}$.  As in \cite{BBL}, to construct exact solutions we must allow an additional source term in the temperature equation:
\begin{equation}\label{tempc}
\ppt{T} = - \bU\cdot \grad T - w \ppz{T} + \frac{k}{\rho c_p} \ppzz{T} + \Sigma + \Sigma_c.
\end{equation}

\subsection{Review: Specification of geometry and temperature}  We repeat the basic specification of the exact solution from \BKL.  Suppose
\begin{equation}\label{coupledh}
    H(t,r) = H_s(r) + \phi(r) \gamma(t)
\end{equation}
where $H_s(r)$ is given by
\begin{align}
    H_s(r) &= \frac{H_0}{(1-1/n)^{n/(2n+2)}} \Big[(1+1/n)s - 1/n + \left(1-s\right)^{1+1/n} - s^{1+1/n}\Big]^{n/(2n+2)}, \label{Hsteady}
\end{align}
and $f(r)$, $g(t)$ are given by
\begin{equation}
\phi(r) = \cos^2\left(\frac{\pi(r- 0.6 L)}{0.6 L}\right)
\end{equation}
if $0.3 L < r < 0.9 L$ and $\phi(r)=0$ otherwise.  Also
\begin{equation}\label{thickperturb}
\gamma(t) = A_p \sin(2\pi t/t_p).
\end{equation}
We suppose from here on that the size of the perturbation $f(r)g(t)$ is limited so that the slope $\partial H/\partial r$ in the positive radial direction is negative.

Now suppose
\begin{equation}\label{coupledT}
    T(t,r,z) = T_s(r) \frac{\nu(t,r)+H(t,r)}{\nu(t,r)+z}
\end{equation}
where $\nu$ is found from $H$ and $T_s$ by
\begin{equation}\label{couplednu}
     \nu(t,r)=\frac{k T_s(r)}{2G} \left(1+\sqrt{1+4\frac{H(t,r) G}{kT_s(r)}}\right).
\end{equation}
Then $T$ satisfies the boundary conditions
    $$T\big|_{z=H} = T_s(r) \quad \text{ and } \quad \frac{\partial T}{\partial z}\Big|_{z=0}=-G/k$$
The form of $T$ in equation \eqref{coupledT} is primarily constrained by the need to do the integral for $\bU$ analytically (below).

\subsection{Computation of the velocity field}  The content presented so far has appeared in \cite{BBL}, but now we start the more detailed computation.

The horizontal velocity is found from equations (3), (4), and (5):
\begin{align}
       \bU(t,r,z) &= - 2 (\rho g)^n A |\grad H|^{n-1} \grad H \int_0^z \exp\left(\frac{-Q}{R T(\zeta)}\right) (H-\zeta)^n\,d\zeta \notag\\
    &= - \br 2 (\rho g)^n A |H_r|^{n-1} H_r e^{-\mu\nu} \int_0^z e^{-\mu\zeta} (H-\zeta)^n d\zeta \notag\\
    &= - \br 2 (\rho g)^n A |H_r|^{n-1} H_r e^{-Q/(RT_s)} \mu^{-(n+1)} \int_{\mu (H-z)}^{\mu H} e^\theta \theta^n\,d\theta, \label{Uchanged}
\end{align}
where $H_r=\partial H/\partial r$ and after a change of variables in the integral.  We have generally suppressed dependence on $t$ and $r$ in these and remaining equations; dependence on $z$ or $\zeta$ will remain explicit.  Here we have also introduced
    $$\mu(t,r) := \frac{Q}{R T_s (\nu + H)}.$$
The change of variables mentioned above is $\theta:=\mu(h-\zeta)$.  It allows us to rewrite
\begin{equation}\label{changevar}
    \int_0^z e^{-\mu\zeta} (H-\zeta)^n\,d\zeta = \mu^{-(n+1)} e^{-\mu H} \int_{\mu(H-z)}^{\mu H} e^\theta \theta^n\,d\theta.
\end{equation}
Note $\bU\big|_{z=0}=0$.

Suppose $n=1,2,3,\dots$ is an integer.  Then we can do integral in \eqref{Uchanged} analytically.  In fact, let $p_n$ be the polynomial defined by the relation
    $$\int e^x\, x^n\,dx = p_n(x)\, e^x + c.$$
It is easy to see by integration-by-parts that $p_n(x)=x^n-n p_{n-1}(x)$ and $p_0(x)=1$.  In particular, it follows that $p_3(x) = x^3-3x^2+6x-6$ and $p_4(x)=x^4-4x^3+12x^2-24x+24$.  Also $p_n(0)=(-1)^n\,n!$.  Assuming that $H_r \le 0$ for all $r$, define
    $$\omega(t,r):= 2(\rho g)^n A (-H_r)^n e^{-Q/RT_s} \mu^{-(n+1)}$$
and
\begin{equation}\label{Indefn}
I_n(t,r,z) := \int_{\mu(H-z)}^{\mu H} e^\theta \theta^n\,d\theta.
\end{equation}
Then we see, by the definition of $p_n$, that
    $$I_n(t,r,z)= p_n(\mu H) e^{\mu H} - p_n(\mu(H-z)) e^{\mu(H-z)}$$
and it follows that
    $$\bU(t,r,z) = + \br \omega I_n(z).$$
We now have an analytical expression for $\bU$ involving no integrals.

Regarding the above calculation, we believe that for ice sheet flow the integer cases $n=1,2,3,4$ completely suffice for numerical testing as the range $1.8\le n \le 4$ is the broadest range of exponents in the constitutive relation known to the authors \cite{GoldsbyKohlstedt}.  It may eventually be appropriate to consider noninteger $n$ cases, in which case the ``incomplete gamma function'' enters \cite{AbramowitzStegun}, but we do not see the need presently.

Next we seek analytical expressions for the vertical velocity $w$ and for the term $\Div \bQ$ which appears in equation (1).  Both of these quantities are vertical integrals of the horizontal divergence  of the horizontal velocity $\bU$.

Recall that the divergence of a radial function is $\Div (f(r)\br)=r^{-1} (\partial/\partial r) \left[r f(r)\right]$.  Also, from now on we will liberally use ``$f_r$'' as an abbreviation for $\partial f/\partial r$.

Thus
\begin{equation*}
    \Div \bU = \frac{1}{r} \frac{\partial}{\partial r} \left[r\omega I_n\right] = \frac{1}{r} \omega I_n  + \frac{\partial \omega}{\partial r} I_n + \omega \frac{\partial I_n}{\partial r}.
\end{equation*}
But
    $$\frac{\partial \omega}{\partial r} =  \omega \left[n \frac{H_{rr}}{H_r} + \frac{Q T_s'}{R T_s^2} - (n+1)\frac{\mu_r}{\mu}\right],$$
noting that $T_s$ is a function of $r$ only, and
\begin{equation*}
    \frac{\partial I_n}{\partial r} = \mu^n e^{\mu H} \left[(\mu_r H +\mu H_r) H^n - (\mu_r (H-z) + \mu H_r) e^{-\mu z} (H-z)^n\right].
\end{equation*}
For the last calculation, recall that if $F(x)=\int_{f(x)}^{g(x)} \phi(t)\,dt$ then $F'(x) = g'(x) \phi(g(x)) - f'(x) \phi(f(x))$.

As mentioned, we will integrate $\Div \bU$ vertically.  In order to facilitate this integration, define
    $$\phi(t,r) := \frac{1}{r} + n\frac{H_{rr}}{H_r} + \frac{Q T_s'}{RT_s^2} - (n+1)\frac{\mu_r}{\mu} \quad \text{and} \quad \gamma(t,r):=\mu^n e^{\mu H} (\mu_r H + \mu H_r) H^n$$
so that
    $$\Div \bU = \omega \left[\phi I_n(z) - \left(\mu^n e^{\mu H} \mu_r\right) e^{-\mu z} (H-z)^{n+1} - \left(\mu^{n+1}e^{\mu H} H_r\right) e^{-\mu z} (H-z)^n + \gamma\right].$$

It follows that
\begin{align}
w(t,r,z) &= -\int_0^z \Div \bU\,d\zeta = \omega \left[\mu^{-1}\left(\mu^{-1}\mu_r-\phi\right) I_{n+1}(z) + \left(\phi(H-z)+H_r\right) I_n(z) - \gamma z\right]. \label{wfinal}
\end{align}
We have again used the change of variable \eqref{changevar}.  Also, we have integrated $I_n$ by changing order of integration:
\begin{align*}
\int_0^z I_n(\zeta)\,d\zeta &= \int_0^z \int_{\mu(H-\zeta)}^{\mu H} e^\theta \theta^n\,d\theta\,d\zeta = \int_{\mu(H-z)}^{\mu H} e^\theta \theta^n \Big(\int_{(H-\theta/\mu)}^z \,d\zeta\Big)\,d\theta \\
    &=(z-H) I_n(z) + \mu^{-1}I_{n+1}(z).
\end{align*}
Note $w\big|_{z=0}=0$.

\subsection{Computation of the compensatory accumulation and heating term}  Next,
\begin{align*}
\Div \bQ &= \grad H \cdot \bU\big|_{z=H} + \int_0^H \Div \bU \,d\zeta = \grad H \cdot \bU\big|_{z=H} - w\big|_{z=H}
\end{align*}
from equation (6) and by the above-mentioned rule for differentiating integrals.  (This is the calculation which shows the equivalence of the vertically-integrated equation of continuity (1) and the surface kinematical equation (10).)  Thus
\begin{align}
\Div \bQ &= - \omega \mu^{-1}\left(\mu^{-1}\mu_r-\phi\right) I_{n+1}(H) + \omega \gamma H. \label{DivQfinal}
\end{align}
We can now compute the compensatory accumulation by using equations \eqref{DivQfinal} and \eqref{coupledh}:
     $$M = \frac{\partial H}{\partial t} + \Div \bQ.$$

Next we get to the strain heating term.  From equations (3), (4), and (5),
    $$\frac{\partial \bU}{\partial z} = -2(\rho g)^n A  \exp\left(-\frac{Q(\nu+z)}{R T_s (\nu + H)}\right) |\grad H|^{n-1} \grad H (H-z)^{n+1}.$$
Thus, from equations (3) and (8),
    $$\Sigma = \frac{1}{\rho c_p} \left(-\rho g (H-z) \grad H\right) \cdot \frac{\partial \bU}{\partial z} = \frac{2(\rho g)^n A\,g}{c_p} \exp\left(-\frac{Q(\nu+z)}{R T_s (\nu + H)}\right) \left(-H_r(H-z)\right)^{n+1}.$$

The above completes much of the hard work.  Recalling equations (1) and (17), we find the desired compensatory heat source:
    $$\Sigma_c = \frac{\partial T}{\partial t} + \bU \cdot \grad T + w \frac{\partial T}{\partial z} - \frac{k}{\rho c_p} \frac{\partial^2 T}{\partial z^2} - \Sigma.$$
It remains to use the chosen functions $H$, $T$ in \eqref{coupledh}, \eqref{coupledT}, respectively, and find the derivatives of $H$ and $T$ which appear in the relevant PDEs.  In particular,
\begin{align*}
     H_t &= f g',\\
     T_t &= T_s \frac{(\nu_t+H_t)(\nu+z)-(\nu+H)\nu_t}{(\nu+z)^2},\\
     T_r &= T_s' \frac{\nu+H}{\nu+z} + T_s \frac{(\nu_r+H_r)(\nu+z)-(\nu+H)\nu_r}{(\nu+z)^2},\\
     T_z &= - T_s \frac{\nu+H}{(\nu+z)^2},\\
  T_{zz} &= 2 T_s \frac{\nu+H}{(\nu+z)^3}.\end{align*}
where ``$F_t$,'' ``$F_r$,'' ``$F_z$,'' ``$F_{zz}$'' denote partial derivatives with respect to the given variables.  (Recall that, however, ``$T_s$'' denotes the surface boundary value for temperature.  Note also that $f,g,T_s$ are functions of a single variable.)

From the above analysis, we see that one must compute at least the following list of analytical derivatives: $H_s', H_s'', f', f'', g', T_s', \nu_t, \nu_r, H_t, H_r, H_{rr}, \mu_r, T_t, T_r, T_z, T_{zz}$.  This is actually done in the reference implementation in the next section.

\section{Reference implementations of Tests \textbf{F} and \textbf{G}}

This section contains a C code which accepts $t,r,z$ and computes $H$, $M$, $T$, $U$, $w$, $\Sigma$, and $\Sigma_c$ for the exact solution in each of Tests \textbf{F} and \textbf{G}.  Both these C codes (and corresponding Fortran 90 codes) are in the public domain under the GNU Public License and are available at
\begin{center}\texttt{www.dms.uaf.edu/$\sim$bueler/iceflowpage.htm}\end{center}
The code here is the authoritative, detailed description of the exact solution.  In particular, it includes the constants used in \BKL, and has analytically-expanded forms for all of the derivatives listed above.  These codes have been compiled using the GNU \texttt{gcc} compiler.

The code is not particularly written for efficiency and can undoubtedly be modified for speed, though perhaps at some loss in clarity.  An efficiency \emph{is} included, namely, the code allows a one-dimensional array of vertical coordinates \texttt{z} as input and returns corresponding arrays for all of the $z$-dependent quantities.  One may therefore ask the subroutine for values of $T$ or $\Sigma_c$ at every depth in a particular column of ice, for instance.  A very simple example program, which merely evaluates the exact solution and prints the result to the standard output, is included below.

\subsection{exactTestsFG.c}  Here is the actual code to compute the exact solutions.

%\Vfile{exactTestsFG_FRAG.c}
\scriptsize\begin{quote}\rule{4.6in}{0.1mm}
\begin{verbatim}
#include <stdio.h>
#include <stdlib.h>
#include <math.h>

double p3(double x) {
  /* p_3=x^3-3*x^2+6*x-6, using Horner's */
  return -6.0 + x*(6.0 + x*(-3.0 + x));
}

double p4(double x) {
  /* p_4=x^4-4*x^3+12*x^2-24*x+24, using Horner's */
  return 24.0 + x*(-24.0 + x*(12.0 + x*(-4.0 + x)));
}

int bothexact(double t, double r, double *z, int Mz,
              double Cp, double *H, double *M, double *TT, double *U,
              double *w, double *Sig, double *Sigc) {

/*
int bothexact(const double t, const double r, const double z[], const int Mz,
              const double Cp, double &H, double &M, double TT[], double U[],
              double w[], double Sig[], double Sigc[]) {
*/

  const double pi = 3.14159265358979;
  const double SperA=31556926.0;  /* seconds per year; 365.2422 days */

  /* parameters describing extent of sheet: */
  const double H0=3000.0;    /* m */
  const double L=750000.0;   /* m */
  /* period of perturbation; inactive in Test F: */
  const double Tp=2000.0*SperA;  /* s */

  /* fundamental physical constants */
  const double g=9.81;       /* m/s^2; accel of gravity */
  const double Rgas=8.314;   /* J/(mol K) */
  /* ice properties; parameters which appear in constitutive relation: */
  const double rho=910.0;    /* kg/m^3; density */
  const double k=2.1;        /* J/m K s; thermal conductivity */
  const double cpheat=2009.0;/* J/kg K; specific heat capacity */
  const double n=3.0;        /* Glen exponent */
  /* next two are EISMINT II values; Paterson-Budd for T<263 */
  const double A=3.615E-13;  /* Pa^-3 s^-1 */
  const double Q=6.0E4;      /* J/mol */
  /* EISMINT II temperature boundary condition (Experiment F): */
  const double Ggeo=0.042;   /* J/m^2 s; geo. heat flux */
  const double ST=1.67E-5;   /* K m^-1 */
  const double Tmin=223.15;  /* K */
  const double Kcond=k/(rho*cpheat);  /* constant in temp eqn */

  /* declare all temporary quantities; computed in blocks below */
  double power, Hconst, s, lamhat, f, goft, Ts, nusqrt, nu;
  double lamhatr, fr, Hr, mu, surfArr, Uconst, omega;
  double Sigmu, lamhatrr, frr, Hrr, Tsr, nur, mur, phi, gam;
  double I4H, divQ, Ht, nut;
  double I4,dTt,Tr,Tz,Tzz;
  int i;
  double *I3;

  I3 = (double *) malloc(Mz * sizeof(double)); /* need temporary array */
  if (I3 == NULL) {
    fprintf(stderr, "bothexact(): couldn't allocate memory\n");
    return -9999;
  }

  if ( (r<=0) || (r>=L) ) {
    printf("\nERROR: code and derivation assume 0<r<L  !\n\n");
    return -9999;
  }

  /* compute H from analytical steady state Hs (Test D) plus perturbation */
  power = n/(2*n+2);
  Hconst = H0/pow(1-1/n,power);
  s = r/L;
  lamhat = (1+1/n)*s - (1/n) + pow(1-s,1+1/n) - pow(s,1+1/n);
  if ((r>0.3*L) && (r<0.9*L))
    f = pow( cos(pi*(r-0.6*L)/(0.6*L)) ,2.0);
  else
    f = 0.0;
  goft = Cp*sin(2.0*pi*t/Tp);
  *H = Hconst*pow(lamhat,power) + goft*f;

  /* compute TT = temperature */
  Ts = Tmin+ST*r;
  nusqrt = sqrt( 1 + (4.0*(*H)*Ggeo)/(k*Ts) );
  nu = ( k*Ts/(2.0*Ggeo) )*( 1 + nusqrt );
  for (i=0; i<Mz; i++)
    TT[i] = Ts * (nu+(*H)) / (nu+z[i]);

  /* compute surface slope and horizontal velocity */
  lamhatr = ((1+1/n)/L)*( 1 - pow(1-s,1/n) - pow(s,1/n) );
  if ( (r>0.3*L) && (r<0.9*L) )
    fr = -(pi/(0.6*L)) * sin(2.0*pi*(r-0.6*L)/(0.6*L));
  else
    fr = 0.0;
  Hr = Hconst * power * pow(lamhat,power-1) * lamhatr + goft*fr;   /* chain rule */
  if ( Hr>0 ) {
    printf("\nERROR: assumes H_r negative for all 0<r<L  !\n");
    return 1;
  }
  mu = Q/(Rgas*Ts*(nu+(*H)));
  surfArr = exp(-Q/(Rgas*Ts));
  Uconst = 2.0 * pow(rho*g,n) * A;
  omega = Uconst * pow(-Hr,n) * surfArr * pow(mu,-n-1);
  for (i=0; i<Mz; i++) {
    I3[i] = p3(mu*(*H)) * exp(mu*(*H)) - p3(mu*((*H)-z[i])) * exp(mu*((*H)-z[i]));
    U[i] = omega * I3[i];
  }

  /* compute strain heating */
  for (i=0; i<Mz; i++) {
    Sigmu = -(Q*(nu+z[i])) / (Rgas*Ts*(nu+(*H)));
    Sig[i] = (Uconst*g/cpheat) * exp(Sigmu) * pow( fabs(Hr)*( (*H) -z[i]) ,n+1);
  }

  /* compute vertical velocity */
  lamhatrr = ((1+1/n) / (n*L*L)) * ( pow(1-s,(1/n)-1) - pow(s,(1/n)-1) );
  if ( (r>0.3*L) && (r<0.9*L) )
    frr = -(2.0*pi*pi/(0.36*L*L)) * cos(2.0*pi*(r-0.6*L)/(0.6*L));
  else
    frr = 0.0;
  Hrr = Hconst*power*(power-1)*pow(lamhat,power-2.0) * pow(lamhatr,2.0)  +
    Hconst*power*pow(lamhat,power-1)*lamhatrr + goft*frr;
  Tsr = ST;
  nur = (k*Tsr/(2.0*Ggeo)) * (1 + nusqrt) +
    (1/Ts) * (Hr*Ts-(*H)*Tsr) / nusqrt;
  mur = ( -Q/(Rgas*Ts*Ts*pow(nu+(*H),2.0)) ) * ( Tsr*(nu+(*H))+Ts*(nur+Hr) );
  phi = 1/r + n*Hrr/Hr + Q*Tsr/(Rgas*Ts*Ts) - (n+1)*mur/mu;   /* division by r */
  gam = pow(mu,n) * exp(mu*(*H)) * (mur*(*H)+mu*Hr) * pow((*H),n);
  for (i=0; i<Mz; i++) {
    I4 = p4(mu*(*H)) * exp(mu*(*H)) - p4(mu*((*H)-z[i])) * exp(mu*((*H)-z[i]));
    w[i] = omega * ((mur/mu - phi)*I4/mu + (phi*((*H)-z[i])+Hr)*I3[i] - gam*z[i]);
  }

  /* compute compensatory accumulation M */
  I4H = p4(mu*(*H)) * exp(mu*(*H)) - 24.0;
  divQ = - omega * (mur/mu - phi) * I4H / mu + omega * gam * (*H);
  Ht = (Cp*2.0*pi/Tp) * cos(2.0*pi*t/Tp) * f;
  *M = Ht + divQ;

  /* compute compensatory heating */
  nut = Ht/nusqrt;
  for (i=0; i<Mz; i++) {
    dTt = Ts * ((nut+Ht)*(nu+z[i])-(nu+(*H))*nut) * pow(nu+z[i],-2.0);
    Tr = Tsr*(nu+(*H))/(nu+z[i])
      + Ts * ((nur+Hr)*(nu+z[i])-(nu+(*H))*nur) * pow(nu+z[i],-2.0);
    Tz = -Ts * (nu+(*H)) * pow(nu+z[i],-2.0);
    Tzz = 2.0 * Ts * (nu+(*H)) * pow(nu+z[i],-3.0);
    Sigc[i] = dTt + U[i]*Tr + w[i]*Tz - Kcond*Tzz - Sig[i];
  }

  free(I3);
  return 0;
}
\end{verbatim}
\rule{4.6in}{0.1mm}\end{quote}\normalsize

\subsection{simpleFG.c}  This is the simple program to exercise the above code.

%\Vfile{simpleFG_FRAG.c}
\scriptsize\begin{quote}\rule{4.6in}{0.1mm}
\begin{verbatim}
#include <stdio.h>
#include <stdlib.h>
#include "exactTestsFG.h"

int main() {

  const double SperA=31556926.0;  // seconds per year; 365.2422 days
  const double Cp=200.0;     // m;  magnitude of the perturbation in test G
  double year, r, HF, MF, HG, MG;
  double *z, *TF, *UF, *wF, *SigF, *SigcF, *TG, *UG, *wG, *SigG, *SigcG;
  double *mb; /* a block of memory */
  int j, Mz;

  printf("Enter  t and r  separated by newline");
  printf(" (in yrs and km, resp.; e.g. 500 500):\n");
  scanf("%lf",&year);
  scanf("%lf",&r);
  printf("Enter  z  values sep by newline (in m);");
  printf(" '-1' to end; e.g. 0 100 500 1500 -1:\n");

  z = (double *) malloc(501 * sizeof(double));
  if (z == NULL) {
    fprintf(stderr, "simpleFG: couldn't allocate memory\n"); return -9999; }

  j=0;
  do {
    scanf("%lf",&z[j]);
    j++;
    if (j>490) printf("\n\n\nWARNING: enter -1 to stop soon!!!\n");
  } while (z[j-1]>=0.0);
  Mz=j-1;

  mb = (double *) malloc(10 * Mz * sizeof(double));
  if (mb == NULL) {
    fprintf(stderr, "simpleFG: couldn't allocate memory\n"); return -9999; }
  TF=mb; UF=mb+Mz*sizeof(double); wF=mb+2*Mz*sizeof(double);
  SigF=mb+3*Mz*sizeof(double); SigcF=mb+4*Mz*sizeof(double);
  TG=mb+5*Mz*sizeof(double); UG=mb+6*Mz*sizeof(double);
  wG=mb+7*Mz*sizeof(double);
  SigG=mb+8*Mz*sizeof(double); SigcG=mb+9*Mz*sizeof(double);

  /* evaluate tests F and G */
  bothexact(0.0,r*1000.0,z,Mz,0.0,&HF,&MF,TF,UF,wF,SigF,SigcF);
  bothexact(year*SperA,r*1000.0,z,Mz,Cp,&HG,&MG,TG,UG,wG,SigG,SigcG);

  printf("\nResults:\n           Test F                         Test G\n");
  printf("(functions of r (resp. t and r) only):\n");
  printf("      H    = %12.6f (m)        H    = %12.6f (m)\n",HF,HG);
  printf("      M    = %12.6f (m/a)      M    = %12.6f (m/a)\n",
         MF*SperA,MG*SperA);
  for (j=0; j<Mz; j++) {
    printf("(z=%10.3f):\n",z[j]);
    printf("      T    = %12.6f (K)        T    = %12.6f (K)\n",TF[j],TG[j]);
    printf("      U    = %12.6f (m/a)      U    = %12.6f (m/a)\n",UF[j]*SperA,
           UG[j]*SperA);
    printf("      w    = %12.6f (m/a)      w    = %12.6f (m/a)\n",wF[j]*SperA,
           wG[j]*SperA);
    printf("      Sig  = %12.6f (*)        Sig  = %12.6f (*)\n",
           SigF[j]*SperA*1000.0,SigG[j]*SperA*1000.0);
    printf("      Sigc = %12.6f (*)        Sigc = %12.6f (*)\n",
           SigcF[j]*SperA*1000.0,SigcG[j]*SperA*1000.0);
  }
  printf("(units: (*) = 10^-3 K/a)\n");

  free(mb);
  return 0;
}
\end{verbatim}
\rule{4.6in}{0.1mm}\end{quote}\normalsize

A run of \texttt{simpleFG} looks like this:

\scriptsize
\begin{quote}\begin{verbatim}
$ simpleFG
Enter  t and r  separated by newline (in yrs and km, resp.; e.g. 500 500):
500
500
Enter  z  values sep by newline (in m); '-1' to end; e.g. 0 100 500 1500 -1:
0
100
500
1500
-1

Results:
           Test F                         Test G
(functions of r (resp. t and r) only):
      H    =  1925.295290 (m)        H    =  2101.899734 (m)
      M    =    -0.010510 (m/a)      M    =     0.040738 (m/a)
(z=     0.000):
      T    =   265.122620 (K)        T    =   267.835036 (K)
      U    =     0.000000 (m/a)      U    =     0.000000 (m/a)
      w    =     0.000000 (m/a)      w    =     0.000000 (m/a)
      Sig  =     0.264346 (*)        Sig  =     1.215392 (*)
      Sigc =    -0.373726 (*)        Sigc =    -1.323664 (*)
(z=   100.000):
      T    =   263.137595 (K)        T    =   265.849860 (K)
      U    =     0.661716 (m/a)      U    =     2.244496 (m/a)
      w    =     0.000005 (m/a)      w    =    -0.000758 (m/a)
      Sig  =     0.173915 (*)        Sig  =     0.817817 (*)
      Sigc =    -0.306255 (*)        Sigc =    -1.022931 (*)
(z=   500.000):
      T    =   255.486095 (K)        T    =   258.194962 (K)
      U    =     1.785938 (m/a)      U    =     6.217140 (m/a)
      w    =     0.000291 (m/a)      w    =    -0.011984 (m/a)
      Sig  =     0.028439 (*)        Sig  =     0.149934 (*)
      Sigc =    -0.199905 (*)        Sigc =    -0.340039 (*)
(z=  1500.000):
      T    =   238.172200 (K)        T    =   240.856843 (K)
      U    =     2.036372 (m/a)      U    =     7.227603 (m/a)
      w    =     0.002288 (m/a)      w    =    -0.050018 (m/a)
      Sig  =     0.000029 (*)        Sig  =     0.000400 (*)
      Sigc =    -0.193301 (*)        Sigc =     0.365908 (*)
(units: (*) = 10^-3 K/a)\end{verbatim}
\end{quote}\normalsize
These numbers allow an easy check on correctness if modifications are made to the implementation exact solutions or, for example, upon recompilation on a new machine.  The numbers can be generally compared to the figures in \cite{BBL}.

\section{Stability and convergence of a numerical scheme for temperature}

\subsection{A traditional finite difference error analysis}  In this section we analyze the error in a semi-implicit, first-order-upwinded finite difference scheme for equation \eqref{temp}, namely the scheme used to generate the results in \BKL.  Appendix A of that paper provides a description of the \emph{coupled} numerical scheme which solves equations \eqref{flow} and \eqref{temp}.  This section ``fleshes out'' Appendix B of that paper, which sketches the error analysis here.

The most important caveats about the analysis here is that \emph{the components of the velocity field are assumed to be known functions independent of the temperature $T$} and also that \emph{the geometry of the ice sheet is assumed fixed}.  In these ways we are not analyzing the coupled numerical scheme.  Nonetheless we believe this analysis provides enough information to help build a reliable adaptive time-stepping scheme and also reveals a significant point of error growth in thermocoupled circumstances.

One can compare the material here to \cite{CDV02} which does a finite element analysis of a moving geometry and velocity field problem for the temperature equation in a flow line model but for which the thermomechanics are only semi-coupled to the geometry.

We generalize equations \eqref{temp} and \eqref{tempc} slightly.  In particular we analyze the equation
\begin{equation}\label{gentemp}
T_t + u(x,y,z,t) T_x + v(x,y,z,t) T_y + w(x,y,z,t) T_z = K T_{zz} + f(x,y,z,t,T),
\end{equation}
for absolute temperature $T$.  We denote the exact solution to \eqref{gentemp} by ``$T$'' or ``$T(x,y,z,t)$.''  The function $f$ in \eqref{gentemp} generalizes $\Sigma$ and $\Sigma+\Sigma_C$ which appear in \eqref{temp} and \eqref{tempc}, respectively.  Note that $\Sigma$ depends upon the temperature, the effective shear stress, and the strain rates.   Thus $\Sigma$ is a function of variables $x,y,z,t,T$ because the velocity components are assumed to be known functions.

We suppose \eqref{gentemp} applies on some bounded region $\Omega\subset \RR^3$ which is fixed in time, and we assume a rectangular computational domain enclosing $\Omega$.  Consider a regular, rectangular grid on that computational domain, in four variables $(x,y,z,t)$, with spacing $\Delta x,\Delta y, \Delta z,\Delta t$ and grid points denoted $(x_i,y_j,z_k,t_l)$.  Let $u_{ijkl}=u(x_i,y_j,z_k,t_l)$, etc, and assume $T_{ijk}^l$ is our numerical approximation of $T(x_i,y_j,z_k,t_l)$.

The numerical scheme is
\newcommand{\Up}[2]{\operatorname{Up}\left(#1\Big|#2\right)}
\begin{align}
\frac{T_{ijk}^{l+1}-T_{ijk}^l}{\Delta t} &+ \frac{\Up{T_{\bullet jk}^l}{u_{ijkl}}}{\Delta x} + \frac{\Up{T_{i\bullet k}^l}{v_{ijkl}}}{\Delta y} + \frac{\Up{T_{ij\bullet}^l}{w_{ijkl}}}{\Delta z} \label{scheme}\\
&\qquad = K\frac{T_{ij,k+1}^{l+1}-2T_{ijk}^{l+1}+T_{ij,k-1}^{l+1}}{\Delta z^2} + f(x_i,y_j,z_k,t_l,T_{ijk}^l) \notag \end{align}
where
    $$\Up{\vf_{\bullet}}{\alpha_i} = \begin{cases} \alpha_i(\vf_i-\vf_{i-1}), & \alpha_i\ge 0 \\ \alpha_i(\vf_{i+1}-\vf_i), & \alpha_i<0.\end{cases}$$
That is, the advection terms are upwinded in the easiest first-order manner and the vertical conduction term is treated implicitly in the easiest centered-difference manner.  Note that we abbreviate the term $f(x_i,y_j,z_k,t_l,T_{ijk}^l)$ by ``$f(T_{ijk}^l)$'' in what follows.

The first kind of error we analyze is local truncation error, but only because it plays a supporting role in the analysis of the approximation error (total numerical error).  The \emph{local truncation error} is defined to be the nonzero result of applying the finite difference scheme to the exact solution \cite{MortonMayers}.  In our case the local truncation error depends upon a long list of cases for the upwind scheme, but listing these turns out to be unnecessary (and certainly uninteresting).  We assume for the next few equations that $u_{ijkl},v_{ijkl},w_{ijkl}$ are all nonnegative, and we will soon be able to return to expressions which apply to all upwinding cases.  (Equation \eqref{scheme} can be rewritten without the ``$\operatorname{Up}(\cdot\big|\cdot)$'' notation if the signs of the velocity coefficients are known, of course.)

Denote the local truncation error at a grid point by $\tau_{ijkl}$, so
\begin{align*}
\tau_{ijkl} &:= \frac{T(x_i,y_j,z_k,t_{l+1})-T(x_i,y_j,z_k,t_l)}{\Delta t} + u_{ijkl} \frac{T(x_i,y_j,z_k,t_l)-T(x_{i-1},y_j,z_k,t_l)}{\Delta x} \\
    &\qquad + v_{ijkl} \frac{T(x_i,y_j,z_k,t_l)-T(x_i,y_{j-1},z_k,t_l)}{\Delta y} + w_{ijkl}\frac{T(x_i,y_j,z_k,t_l)-T(x_i,y_j,z_{k-1},t_l)}{\Delta z} \\
    &\qquad - K\frac{T(x_i,y_j,z_{k+1},t_{l+1}) -2T(x_i,y_j,z_k,t_{l+1})+T(x_i,y_j,z_{k-1},t_{l+1})}{\Delta z^2} \\
    &\qquad - f(x_i,y_j,z_k,t_l,T(x_i,y_j,z_k,t_l)).
\end{align*}
The finite difference quotients here all have well-known Taylor expansions.  Because $T(x,y,z,t)$ solves \eqref{gentemp}, it follows that
    $$\tau_{ijkl} = O(\Delta t,\Delta x,\Delta y,\Delta z).$$
Including the Taylor expansions and all cases of upwinding,
\begin{align}
\tau_{ijkl} &= \frac{1}{2} T_{tt} \Delta t \pm \frac{u_{ijkl}}{2} T_{xx} \Delta x \pm \frac{v_{ijkl}}{2} T_{yy} \Delta y \pm \frac{w_{ijkl}}{2} T_{zz} \Delta z -\frac{K}{12} T_{zzzz} \Delta z^2
\end{align}
where the higher partial derivatives of $T$ are evaluated somewhere in the grid neighborhood of $(x_i,y_j,z_k,t_l)$.  We see that the finite difference scheme is \emph{consistent} \cite{MortonMayers}.  Note that if $w_{ijkl}=0$ then $\tau_{ijkl}=O(\Delta z^2)$, but this case is too special to be of interest.

Let
    $$e_{ijk}^l=T_{ijk}^l - T(x_i,y_j,z_k,t_l)$$
be the (signed) error at a grid point.  Because $T_{ijk}^l$ solves difference scheme \eqref{scheme} exactly, and by the definition of the local truncation error,
\begin{align}
&\frac{e_{ijk}^{l+1}-e_{ijk}^l}{\Delta t} + u_{ijkl} \frac{e_{ijk}^l-e_{i-1,jk}^l}{\Delta x} + v_{ijkl} \frac{e_{ijk}^l-e_{i,j-1,k}^l}{\Delta y} + w_{ijkl}\frac{e_{ijk}^l-e_{ij,k-1}^l}{\Delta z} \label{errorsolves} \\
    &\qquad\qquad - K\frac{e_{ij,k+1}^{l+1} - 2e_{ijk}^{l+1} + e_{ij,k-1}^{l+1}}{\Delta z^2} + \left[f(T(x_i,y_j,z_k,t_l))- f(T_{ijk}^l)\right] = - \tau_{ijkl}.  \notag
\end{align}

We are interested in the evolution of error so we solve for the error at grid point $(x_i,y_j,z_k)$ at time $t_{l+1}$:
\begin{align}
\left(1 + 2\frac{K\Delta t}{\Delta z^2}\right) e_{ijk}^{l+1} &= \frac{K\Delta t}{\Delta z^2} \left(e_{ij,k+1}^{l+1} + e_{ij,k-1}^{l+1}\right) + \left\{1-u_{ijkl}\frac{\Delta t}{\Delta x}-v_{ijkl}\frac{\Delta t}{\Delta y}-w_{ijkl}\frac{\Delta t}{\Delta z}\right\} e_{ijk}^l \label{errornext} \\
    &\qquad +u_{ijkl}\frac{\Delta t}{\Delta x}e_{i-1,jk}^l + v_{ijkl}\frac{\Delta t}{\Delta y}e_{i,j-1,k}^l + w_{ijkl}\frac{\Delta t}{\Delta z}e_{ij,k-1}^l \notag \\
    &\qquad - \Delta t\left[f(T(x_i,y_j,z_k,t_l))- f(T_{ijk}^l)\right] - \Delta t\,\tau_{ijkl} \notag
\end{align}
Note that the quantity in curly braces actually varies depending on the upwinding case, as do all coefficients of the errors which depend on the velocities.  The possibilities for the quantity in curly braces are described by
\begin{equation}1\pm u_{ijkl}\frac{\Delta t}{\Delta x} \pm v_{ijkl}\frac{\Delta t}{\Delta y} \pm w_{ijkl}\frac{\Delta t}{\Delta z}, \label{specialterm}
\end{equation}
with, for example, a ``$-$'' in front of $u_{ijkl}$ if $u_{ijkl}>0$ and a ``$+$'' if $u_{ijkl}<0$.

The next step is significant.  We identify an assumption which is \emph{sufficient}, under the noted additional assumptions of smooth and fixed velocity fields and geometry, but which is presumably not necessary, for the stability of our finite difference scheme.  This assumption is part of a maximum principle argument for a finite difference scheme; see standard examples of such arguments in \cite{MortonMayers}.

\begin{ass} \label{ass:CFL} The space-time grid satisfies
\begin{equation}\label{CFLfirst}
    1-|u_{ijkl}|\frac{\Delta t}{\Delta x}-|v_{ijkl}|\frac{\Delta t}{\Delta y}-|w_{ijkl}|\frac{\Delta t}{\Delta z} \ge 0
\end{equation}
or, equivalently, the time step is chosen small enough to satisfy
\begin{equation}\label{CFLsecond}
    \Delta t \le \left(\frac{|u_{ijkl}|}{\Delta x} + \frac{|v_{ijkl}|}{\Delta y} + \frac{|w_{ijkl}|}{\Delta z}\right)^{-1}.
\end{equation}
\end{ass}

As is standard in these maximum principle arguments, the plan now is to use this assumption to conclude that all coefficients of errors are nonnegative in \eqref{errornext}, and thus to bound the error.  Let
    $$E^l = \max_{i,j,k} |e_{ijk}^l| \qquad \text{ and } \qquad \tau_l = \max_{i,j,k} |\tau_{ijkl}|.$$
In particular, $E^l$ is the maximum absolute approximation error over the whole grid at time $t_l$.  It would be nice to know that it does not grow too large; such is our goal.  Equation \eqref{errornext} and all of its upwinding variations collectively imply
\begin{align}
\left(1 + 2\frac{K\Delta t}{\Delta z^2}\right) |e_{ijk}^{l+1}|
 &\le  2 \frac{K\Delta t}{\Delta z^2} E^{l+1} + \left\{1-|u_{ijkl}|\frac{\Delta t}{\Delta x}-|v_{ijkl}|\frac{\Delta t}{\Delta y}-|w_{ijkl}|\frac{\Delta t}{\Delta z}\right\} E^l \label{errorthird} \\
    &\qquad +|u_{ijkl}|\frac{\Delta t}{\Delta x}E^l + |v_{ijkl}|\frac{\Delta t}{\Delta y}E^l + |w_{ijkl}|\frac{\Delta t}{\Delta z} E^l \notag \\
    &\qquad + \Delta t\left|f(T(x_i,y_j,z_k,t_l)) - f(T_{ijk}^l)\right| + \Delta t\,\tau_l. \notag
\end{align}
The reader should note that we have used Assumption \ref{ass:CFL} to derive \eqref{errorthird}.

We have not yet arrived at the desired error inequality.  In fact, the size of the ``$\big|f(T(x_i,y_j,z_k,t_l)) - f(T_{ijk}^l)\big|$'' term is very important, and we must make another assumption which amounts to knowing that the derivative of $f$ with respect to the temperature $T$ is not too big.  Though this assumption is fully justified in the case of the strain-heating term $\Sigma$ for ice, the Lipshitz coefficient which appears (below) is interesting in connection to the ``spokes''.

\begin{ass} \label{ass:Lip}  For the source term $f$ in \eqref{gentemp}, there exists a bounded nonnegative function $L(x,y,z,t)\ge 0$ such that
    $$\left|f(x,y,z,t,T_1)-f(x,y,z,t,T_2)\right| \le L(x,y,z,t) |T_1-T_2|.$$
In particular, if the partial derivative $\partial f/\partial T$ exists and is bounded, and if we define
    $$L_f(x,y,z,t) = \max_{T_{\text{min}} \le T \le T_{\text{max}}} \left|\frac{\partial f}{\partial T}(x,y,z,t,T)\right|,$$
then we may take $L=L_f$ by the Mean Value Theorem.  Here ``$T_{\text{min}}$'' and ``$T_{\text{max}}$'' are lower and upper bounds, respectively, on the exact solution. \end{ass}

Note that an obvious lower bound on the exact solution $T$ is the minimum value of the surface temperature in standard ice sheet circumstances.  An upper bound is the pressure-melting temperature, in practice.  The actual polythermal nature of ice sheets is obviously ignored here; compare \cite{Greve}.

One may call $L$ in Assumption \ref{ass:Lip} a \emph{local Lipshitz constant} for $f$ (as a function of $T$).  Let $L_{ijkl}=L(x_i,y_j,z_k,t_l)$ be its grid value.  Let $\bar L=\sup_{(x,y,z)\in\Omega,t\ge 0} L(x,y,z,t)$ be the bound on $L$, the global Lipshitz constant.

We apply Assumption \ref{ass:Lip} to \eqref{errorthird} and collect terms by noting that several on the right have coefficients which add to one.  Finally we recall the definition of $e_{ijk}^l$.  The result is:
\begin{equation}\label{errgrowthloc}
\left(1 + 2\frac{K\Delta t}{\Delta z^2}\right) |e_{ijk}^{l+1}|
 \le 2 \frac{K\Delta t}{\Delta z^2} E^{l+1} + E^l +  \Delta t\,L_{ijkl} \, |e_{ijk}^l| + \Delta t\,\tau_l.
\end{equation}
This is a significant error inequality.

\subsection{A convergence theorem}  One consequence of inequality \eqref{errgrowthloc} is the following convergence theorem.

\begin{thm} Assume that the velocity components $u,v,w$ are smooth.  Assume that the geometry is fixed and is sufficiently smooth (so that the local truncation error $\tau_{ijkl}$ is indeed $O(\Delta t)$).  If Assumptions \ref{ass:CFL} and \ref{ass:Lip} apply then the error grows at an (at most) exponential rate in the total model run time $t_f$, times an $O(\Delta t)$ factor:
    $$E^l \le t_f\, \exp(t_f\,\bar L)\, \left(\max_{0 \le k \le l-1} \tau_k\right) \le t_f \exp(t_f\,\bar L)\, O(\Delta t).$$
Therefore the scheme converges as the time step $\Delta t$ goes to zero.
\end{thm}

\begin{proof}  By taking maximums over the grid, inequality \eqref{errgrowthloc} implies
    $$E^{l+1}  \le  \left(1 + \Delta t\,\bar L\right) \, E^l + \Delta t\,\tau_l.$$
We initialize the scheme with the correct initial values so $E^0=0$.  By induction, therefore,
    $$E^l \le \Delta t \left[\left(1 + \Delta t\,\bar L\right)^l  \tau_0 + \left(1 + \Delta t\,\bar L\right)^{l-1} \tau_1 + \dots + \left(1 + \Delta t\,\bar L\right)^1 \tau_{l-2} +  \tau_{l-1}\right].$$
Recall that $(1+x/n)^n \le \exp x$ for $x\ge 0$ and $n\ge 0$.  It follows that
\begin{align*}
E^l &\le \Delta t \left(\max_k \tau_k\right) \left[\left(1 + \Delta t\,\bar L\right)^l + \left(1 + \Delta t\,\bar L\right)^{l-1} + \dots + \left(1 + \Delta t\,\bar L\right)^1 + 1\right] \\
    &\le \Delta t \left(\max_k \tau_k\right) l\, \left(1 + \Delta t\,\bar L\right)^l \le l \Delta t \left(\max_k |\tau|_k\right) \left(1 + \frac{t_f \bar L}{N} \right)^N \\
    & \le l \Delta t \left(\max_k \tau_k\right) \exp(t_f \bar L) \le t_f\,\exp(t_f\bar L) \left(\max_k \tau_k\right)
\end{align*}
as claimed.  (Note that the total time $t_f$ is the number of time steps multiplied by the step size: $t_f = N \Delta t$.  In particular, for each step $l$ we have $l\Delta t \le N \Delta t = t_f$.)\end{proof}

This theorem is desirable but it is not directly useful.  It is not surprising given the strong smoothness assumptions and it is not practical because the convergence constant ``$t_f \exp(t_f\,L)$'' is almost certainly too large in practice.

Dismissing this theorem for now, we see three intermediate results of importance, namely the two Assumptions, which arise reasonably naturally, and the inequality \eqref{errgrowthloc}.

\subsection{Adaptive time-stepping}  Note that inequality \eqref{CFLsecond} is used in the adaptive time-stepping numerical scheme described in \BKL.  By default the mass-balance and temperature steps are synchronized.  In this synchronized case the stability condition for the mass-balance scheme---see \BKL---and constraint \eqref{CFLsecond} are combined to give the time step
    $$\Delta t= \min\left\{0.12 \times 2 \left(\frac{1}{\Delta x^2}+\frac{1}{\Delta y^2}\right)^{-1}\,\Big(\max D_{ijl}\Big)^{-1} \,,\quad \max \left(\frac{|u_{ijkl}|}{\Delta x} + \frac{|v_{ijkl}|}{\Delta y} + \frac{|w_{ijkl}|}{\Delta z}\right)^{-1}\right\}$$
where $D_{ijl}$ is the computed value of the diffusivity of the mass-balance equation (\BKL).  The special value ``$0.12$'' is essentially empirical, i.e.~it results from testing in a variety of circumstances.  The ``$\max$''s which appear in the above equation are computed over the spatial grid at time $t_l$.

For fixed velocity and diffusivity fields, and if $\Delta x=\Delta y$, we see that $\Delta t = O(\Delta x^2)$ as $\Delta x \to 0$ because of the mass-balance condition.  As the grid is refined it might be expected that the mass-balance constraint will inevitably become active, but in fact the maximum magnitude of $w_{ijkl}$ near the margin causes constraint \eqref{CFLsecond} to be active in many cases including some EISMINT II simulations and when using a digital elevation map for the surface elevation of Antarctica.  (See \BKL on the source of large vertical velocities.)  For the better behaved exact Tests \textbf{F} and \textbf{G}, and for realistic ice sheet simulations wherein the geometry has been smoothed by evolution and the grid is significantly refined, we do indeed see the mass-balance condition most active.

One can reasonably consider a scheme which tolerates violations by a modest factor of the CFL constraint \eqref{CFLsecond} at a few grid points.  The cost would be possible loss of accuracy at locations where the computation is not likely to be accurate anyway (e.g.~at margins, near mountain ranges within realistic ice sheets, or at transitions from frozen to sliding base if that transition is too abrupt anyway).  The benefit would be that the time step for the whole sheet could be longer and computation times could be reduced.

\subsection{Emergence of spokes}  Let us reconsider error inequality \eqref{errgrowthloc}.  Suppose that we look at the worst case location on the grid at time $t_{l+1}$, that is, at $i,j,k$ such that $|e_{ijk}^{l+1}|=E^{l+1}$.  In that case \eqref{errgrowthloc} says
\begin{equation}\label{errlocworst}
|e_{ijk}^{l+1}| \le E^l +  \Delta t\,L_{ijkl} \, |e_{ijk}^l| + \Delta t\,\tau_l \qquad \qquad (\text{\emph{applies at worst case grid point}}).
\end{equation}
This inequality says that the error at the worst-case grid point can be no larger than the worst error over the grid at the previous time step plus two terms which are proportional to the time step.  It is a statement of limited growth of error \emph{unless} these two terms happen to be big at time $t_l$.

The first of these terms depends on the local size of the strain heating term, or, more precisely, its variability with respect to temperature.  The second, involving the maximum local truncation error over the grid, depends on the smoothness of the exact solution, or, more precisely, on the degree to which it does not satisfy the finite difference scheme.  We believe that because of the free boundary nature of real ice sheet problems, the second truncation error term is in fact likely to be large at points.  It causes only arithmetic growth in error, however, and it depends purely on the smoothness of the exact solution rather than nonlinear and nontrivially evolving errors on the grid.  In any case we concentrate on the first term because we believe it is involved in the emergence of the spokes in the EISMINT II results.

It is an easy calculation that if a flow law of form \eqref{ArrGlenNye} is used then
\begin{equation}\label{dfdT}
    \left|\ppT{f}\right| = \left|\ppT{\Sigma}\right| = \frac{2 A Q}{\rho c_p R}\, \sigma^{n+1} \exp\left(-\frac{Q}{R T}\right) \frac{1}{T^2} = \Sigma \, \frac{Q}{R T^2}.
\end{equation}
Thus we may use
    $$L_{ijkl} = \Sigma_{ijkl} \, \frac{Q}{R T_{ijkl}^2}$$
for the local Lipshitz constant in \eqref{errlocworst}.

If, in fact, the two-regime Paterson-Budd (1982)\nocite{PatersonBudd} flow law is used then we need the \emph{larger} activation energy for warmer ice, $Q_{T\ge 263.15} = 13.9 \times 10^4 \,\text{J}\,\text{mol}^{-1}$, in our worst case estimate of the local Lipshitz constant, rather than the cold value $Q_{T< 263.15} = 6.0 \times 10^4 \,\text{J}\,\text{mol}^{-1}$.  Thus the rate of error growth jumps up as one moves from relatively cold ice to warmer ice, precisely into the region of warm spokes in EISMINT II experiment F \cite{EISMINT00}.  (The Hooke (1981)\nocite{Hooke} relation should behave in roughly the same manner.  Note that Payne and Bladwin (2000)\nocite{PayneBaldwin} produced spokes for EISMINT II experiment F using both the Paterson-Budd and Hooke relations, though the details of the spokes differ.)  In any case the quantity $\left|\partial\Sigma/\partial T\right|$ has a characteristic spatial variation, illustrated in \BKL, which strongly suggests that locations where this quantity is large are locations of the emergence of (warm) spokes.

%         References
%\bibliography{../ice}
%\bibliographystyle{agsm}

\end{document}